# The Tiny Triplet Finder as a Versatile Track Segment Seeding Engine for Trigger Systems

Jinyuan Wu

***Abstract*— In high energy physics experiment trigger systems, track segment seeding is a resource consuming function and the primary reason is the computing complexity of the segment finding process. As the Moore's' Law is reaching its physical limit, reducing computing complexity should be carefully considered, rather than keep piling up silicon resources. The Tiny Triplet Finder is a scheme that reduces the computing complexity of the segment seeding. As a proof of concept, a 3D track segment seeding engine core based on the Tiny Triplet Finder has been implemented and tested in a low-cost FPGA device. The seeding engine is designed to preselect and group hits (stubs) from detector layers to feed subsequent track fitting stage. The seeding engine consists of a Hough transform space for r-z view and a Tiny Triplet Finder for r-phi view to implement 3D constraints. The seeding engine is organized as a pipeline so that each hit is processed in a single clock cycle. Taking advantage of the register-like storage block scheme which enables effectively resetting of a block RAM within a single clock cycle, clearing or refreshing the seeding engine takes only one clock cycles between two events. The Tiny Triplet Finder is also a generic coincidence finding scheme that can be used for many tasks. As a versatility demonstration, track segment finding performances for two distinctive detector geometries are tested in our seeding engine. In a collider barrel-layer geometry, the fake segment rates are studied for 3D (i.e., both r-phi and r-z views) and 2D (i.e., r-phi or r-z view only) configurations for high hit multiplicity events (>4000 hits/layer in the barrel region). Another detector geometry contains strip plane layers with timing information. The numbers of coincidences, both real or fake, with or without timing ("3D" or "2D") information at various hit multiplicities are studied.**

## I. Introduction

THE Tiny Triplet Finder is a generic coincidence finding scheme that can be used for track segment seeding processes in either trigger systems or higher level event selection stages for various detector geometries. In the Tiny Triplet Finder, hit patterns of several detector layers are shifted simultaneously as inputs to feed a set track segment roads for coincidence finding. With Tiny Triplet Finder, multiple sets of similar track segment coincidence roads are combined into a single set which reduces silicon resource demands significantly. The only requirement to a detector geometry is an approximate shift symmetry which is satisfied for most detector designs today. A block diagram of our seeding engine based on the Tiny Triplet Finder is shown in Fig. 1.

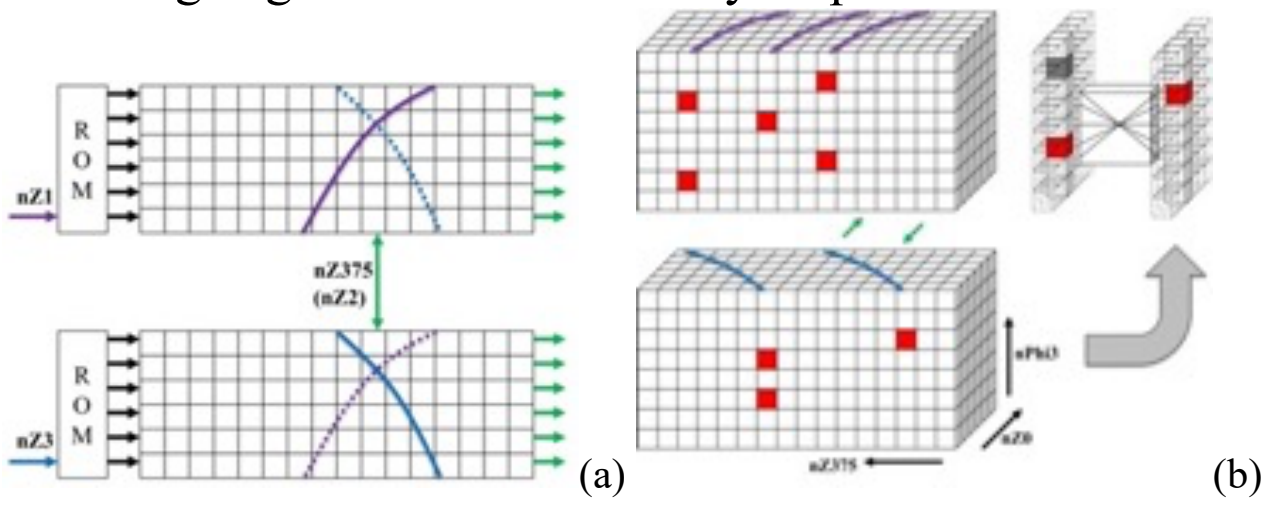

Fig. 1. (a) The hit storage blocks (b) The details storage format and the Tiny Triplet Finder

To verify the versatility of the Tiny Triplet Finder track seeding engine, tracking performances are studied for two different detector geometries with the same seeding engine firmware implemented in a low-cost field-programmable-gate-array (FPGA) device in this work. The first geometry is a collider barrel-layer configuration and the second is a strip plane timing layers configuration. The changes to the firmware include some lookup table contents and coincidence road map which are generated automatically with the simulation software.

The track seeding engine is not intended to perform the entire tracking process, but rather, it is merely a seed feeder to the later track fitter stage (such as a Kalman filter stage). What being fed to the track fitting stage contain sets of hits of track segment candidates, both good tracks and unavoidably many fake tracks. Therefore, a primary requirement for the seeding engine is that the numbers of fake track segments sent to later stage are controlled within the processing capability. For both geometries, total track segment candidates (both good and fake) to be processed by the fitting stage are compared for different configurations (i.e., 3D, 2D or intermediate arrangements) and various event complexities (i.e., numbers of hits per events).

In the collider barrel-layer geometry, the seeding engine serves a 10-degree by 240 cm section with up to 112 hits per layer per events. This hit multiplicity equivalents to >4000 hits per layer per event in the full barrel region. The engine takes 112 clock cycles to fetch the hit data into the internal memories of the engine, another 112 clock cycles to find coincidence of the track segment roads and one clock cycle to refresh the engine for the next event. The test results shows that as expected, the 3D coincidence (in both r-phi and r-z views) yields significantly less fake track segment candidates than 2D configurations (meaning r-phi or r-z view only). With less fake track segment candidates, number of copies of the track fitters in the later stage can be reduced which otherwise would need to be massively duplicated.

In the strip plane timing layer geometry, the detector planes produce both coordinates and arrival time of the hits. The test results show that, as expected, the fake track segment candidates for coincidences with timing information are significantly less than ones without timing information.

## II. The Test Geometries for the 3D Track Segment Seeding Engine

The first test geometry is an idealized collider detector with cylindrical symmetry is chosen and a typical event in it is shown in Fig. 1. The collision points are along the axis and a solenoidal magnetic field is applied. A sector of the detector with 10 degrees in □ and full length (up to 240 cm) in z are handled by a seeding engine core. For each event in the 10-degree sector, 10 tracks with high transverse momentum (> 2GeV/c) along with up to 102 other random hits (from low momentum tracks) per layer are generated, which allows an array of 36 seeding engine cores to handle an event as complex as >4000 hits per layer in the full barrel region. A collision region of ±10 cm along z direction and a vertex displacement range of ±2 mm away from the z-axis are included.

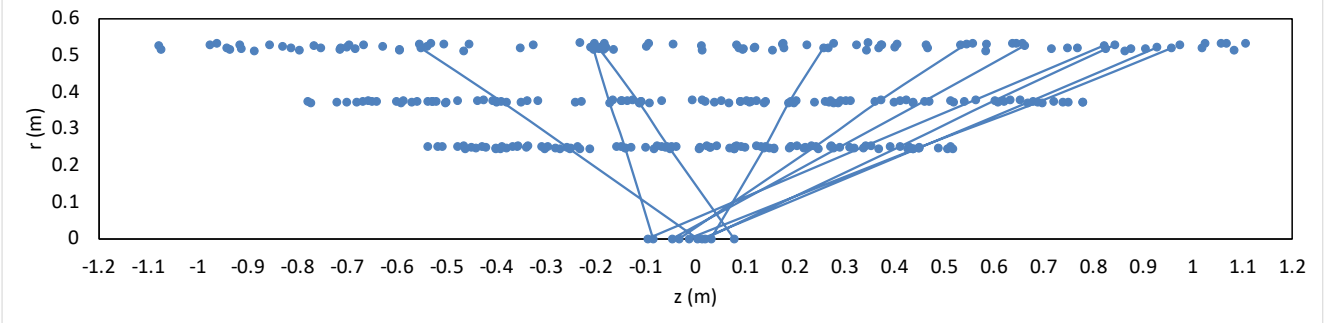

| | |
|---|---|
| Magnetic Field, Minimum PT | 4T, 2 GeV/c |
| Layer 1 | R1: 0.250 (m); L1: 2.4 (m); $\phi$: 10 $\pm$2.2 => 14.3 (deg.) |
| Layer 2 | R2: 0.375 (m); L2: 2.4 (m); $\phi$: 10 (deg.) |
| Layer 3 | R3: 0.525 (m); L3: 2.4 (m); $\phi$: 10 $\pm$2.6 => 15.2 (deg.) |
| Bin size of z375, z1 and z3 | 1 cm to 10 cm, 240 bins to 2 bins over 240 cm |
| Bin size in $\phi$<br>Layer 1<br>Layer 2<br>Layer 3 | 0.125 to 2 (deg.), 128 bins to 8 bins over 16 degrees<br>0.545 (mm)<br>(Both z and $\phi$ coordinate in Layer 2 use full resolution.)<br>1.145 (mm) |
| Bin size of z0 | 2 cm, 10 bins to cover $\pm$ 10 cm |

Fig. 2. A typical event in a model detector with collider geometry

The bin sizes in z and $\phi$ directions shown above are used in the engine

The author is with Fermi National Accelerator Laboratory, Batavia, IL 60510 USA (phone: 630-840-8911; fax: 630-840-2950; e-mail: jywu168@ fnal.gov).

core. The track projection in r-z view is specified with z0 and z375. The parameter z0 is defined as the z coordinate along z-axis while z375 is the one when the track passing a virtual cylinder surface with radius 375 mm. (In fact, z-coordinate of a hit on Layer 2 can be roughly considered as z375).

A strip plane timing layer geometry is shown in Fig. 2 with constraint in both spatial and temporal domains.

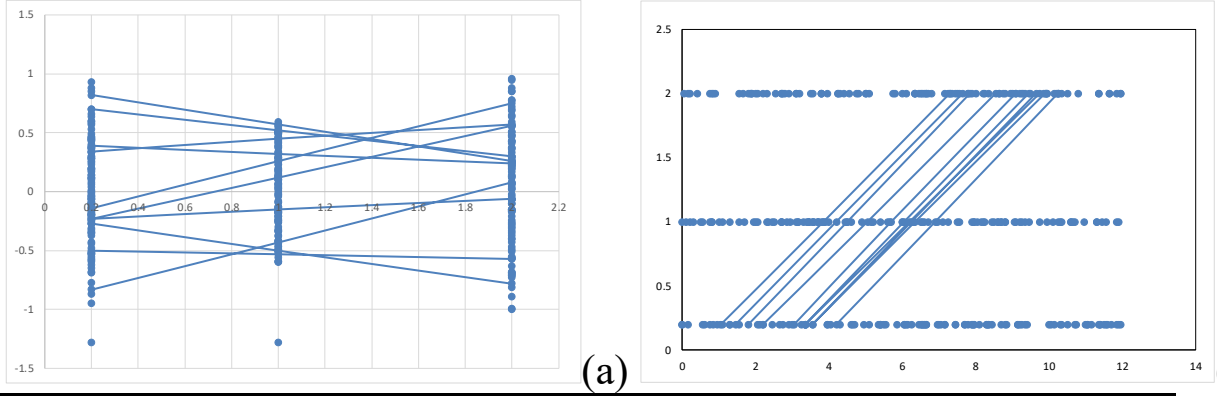
(a) (b)

| YZ view | Non-bend view |
|---|---|
| Layer 1 | z1: 0.2 (m) |
| Layer 2 | z2: 1.0 (m) |
| Layer 3 | z3: 2.0 (m) |
| Bin size of y1 and y3 | 1 cm |
| Bin size in t | 50 ps, 500 ps, 2 ns |

Fig. 3. A typical event in the strip-plane/timing detector

The constraint in spatial domain is in the YZ view demanding hits in Layer 1, 2 and 3 form a straight line as shown in Fig. 3(a). The timing constraint requires that times of flight between these layers are proportional to the spacing of the layers, or in other words, these 3 hits in the t-z view form a "straight line", as shown in Fig. 3(b).

## III. Test Results

The 3D seeding engine core has been implemented in an Altera Cyclone 5 FPGA (5CGXFC5C6F27C7N) and tested in an evaluation module (Terasic C5G) with 250 MHz system clock. The firmware uses approximately 32.9K logic cells, 150 RAM blocks and 100 DSP multipliers. The test station is shown in Fig. 4(a).

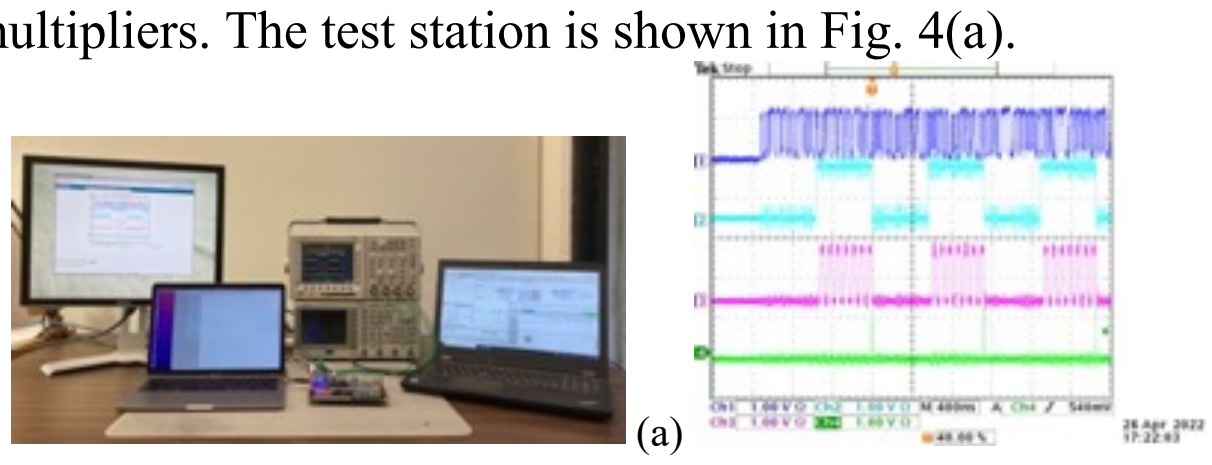
(a) (b)

Fig. 4. The test stand and the oscilloscope screen shot of the operating timing

Simulated hit data are transmitted from a computer via USB cable to the FPGA. The track road coincidence results produced by the seeding core are sent back to the computer and stored in a file for further analysis. Operation timing is shown in Fig. 4(b) in which three events are processed with no gaps between them. CH 1 in the oscilloscope screen shot is a data bit in Layer 1 in the seeding engine. When the data of several events are transmitted to the memory buffer in the FPGA, the engine starts to process events in burst fashion with data in Layer 1 and 3 transmitted into the engine first. Once the Layer 1 and 3 data are transmitted, the engine operated in the second phase as indicated by the CH 2, in which coincidences are searched and sent out to the later stage, which can be seen in CH 3 representing a bit of the write address in the buffer memory. After coincidence searching, a refresh signal is issued as shown in CH 4 above. The refresh signal prepares the entire seeding engine, including its internal memory, for the next event. The next event can be processed immediately after the refresh command.

A processing result of the seeding core under an arbitrary complexity (10 good track plus 102 fake hits per layer within a 10-degree detector sector) is shown above in Fig. 4. Most good tracks (>99%) are accepted as indicated by the red histogram. After the selection of the seeding engine, most of fake hits (~80%) are in the bin 0 meaning that there are no coincidences for them and they are rejected (yellow).

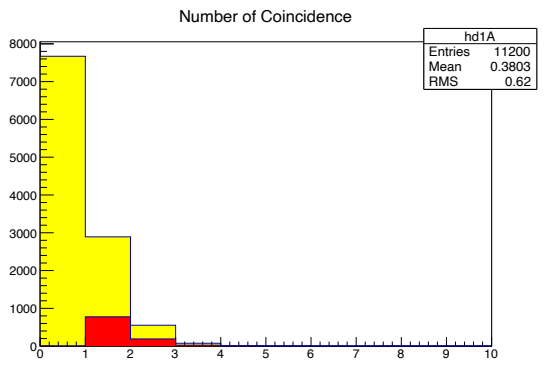

Fig. 4. Number of coincidence for the 3D seeding engine for bin width 0.125 degrees in phi and 1 cm in z

The detector granularity in the engine can be changed relatively flexible by using simulation program to generate new files for the firmware. We have changed the bin width in phi from 0.125 degrees to 0.5 and 2 degrees. Note that when the bin width in phi becomes 2 degrees, the coincidence searching is almost a 2D one with r-z view only. The results are plotted in Fig. 5.

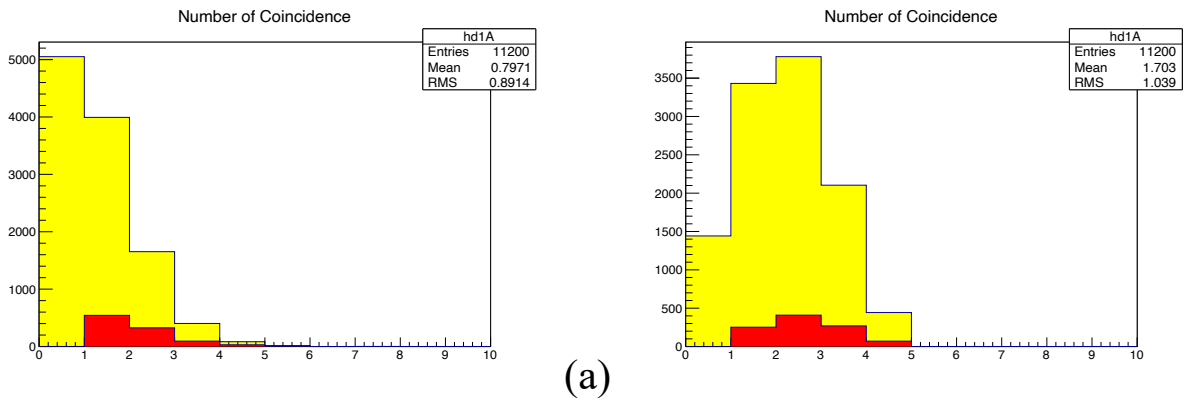

(a) (b)

Fig. 5. Number of coincidence for the 3D seeding engine with different bin width in phi. (a): 0.5 degrees and (b): 2 degrees

It can be seen that the fake hits in bins 1 and above become more meaning that there are many fake combinations must be checked and rejected by the later (Kalman filter) stage. As expected, 2D coincidence only in r-z view has a worse fake rejection capability than the 3D one.

For the strip-plane/timing geometry, we have studied number of segment candidates (both good and fake) that must be sent to the later stage for further checking and rejecting. The simulated events contains 10 good tracks and 10, 30, 50, 70 and 102 random fake hits. The coincidence is a "3D" one including constraints in both y-z view and t-z view. Different timing measurement bin widths, 50 ps, 500 ps and 2 ns are checked in the study to understand the effects of the timing constraint. The test results are plotted in Fig. 6.

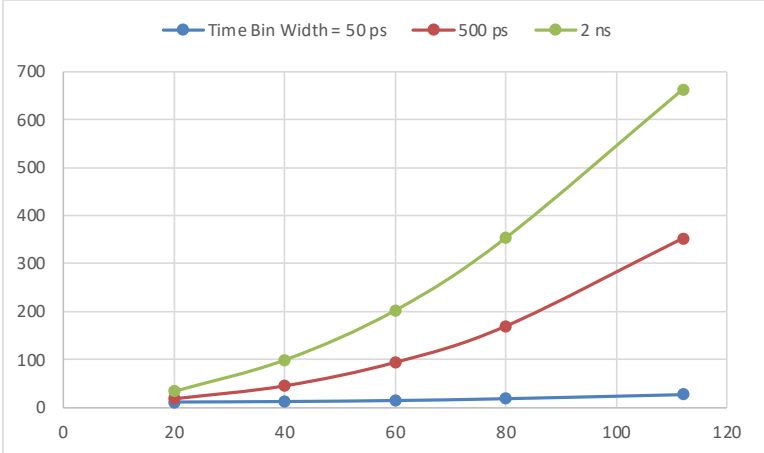

Fig. 6. Number of segment candidates found by the segment seeding engine at different event complexity

Clearly when number of hits per layer per event is not very high, the seeding engine sends almost no fake segment candidates to the later stage. This is true for detectors with or without timing information. When the complexity of the event increases, the segment candidates sent to later stage are dominated with fake ones. However, with good timing constraints, the users are able to eliminate significant amount of the fake candidates which helps to reduce the complexity and cost of the later stage and the entire trigger system.

## IV. Conclusions

A 3D track segment seeding engine core has been implemented in a low-cost FPGA and tested with two detector geometries. The first geometry in this exercise is a 3D segment finding and it is a very difficult seeding task. The primary purpose of this work is to demonstrate the existence and availability of low resource usage approaches. An exercise on the strip-plane/timing geometry demonstrated the versatility of this approach. The building blocks of the register-like storage block and the Tiny Triplet Finder can also be used in other online trigger or offline tracking acceleration projects.

The Tiny Triplet Finder is a scheme that will help to fit the triplet finding algorithm into reasonable sized FPGA. In typical track segment finding implementations, we would have to implement many track roads in coincidence logic which would use large silicon area. In Tiny Triplet Finder, we use shifters to rotate hit patterns to align with the coincidence logic. This way, we will only need to implement one set of the roads. The low silicon area usage helps to reduce costs not only in FPGA devices, but also in power consumption, cooling and printed circuit board complexity, which eventually turns some impossible into possible.